\documentstyle[preprint,prl,aps]{revtex}

\begin{document}
\preprint{}

\title{Constancy of the bilayer splitting as a function of doping in
$Bi_{2}Sr_{2}CaCu_{2}O_{8+\delta}$}
\author{Y. -D. Chuang$^{1,2}$, A. D. Gromko$^{1}$, A.V.
Fedorov$^{1,2}$, Y.
Aiura$^{3}$, K. Oka$^{3}$, Yoichi Ando$^{4}$, D. S. Dessau$^{1}$
}
\address{$^{1}$Department of Physics, University of Colorado, Boulder,
Colorado 80309-0390}
\address{$^{2}$Advanced Light Source (ALS), Berkeley, CA 94720}
\address{$^{3}$National Institute for Advanced Industrial Science and
Technology (AIST)
AIST Tsukuba Central, 1-1-1 Umezono, Tsukuba, Ibaraki 305-8568,Japan}
\address{$^{4}$Central Research Institute of Electric Power Industry
(CRIEPI), 2-11-1 Iwato-Kita, Komae, Tokyo 201-8511, Japan}

\date{Received \today}
\maketitle


\begin{abstract}
Using high energy resolution angle resolved photoemission
spectroscopy, we have resolved the bilayer splitting effect in a wide
range of dopings of the bilayer cuprate
$Bi_{2}Sr_{2}CaCu_{2}O_{8+\delta}$.  This bilayer splitting is due to
a nonvanishing intracell coupling $t_{\perp}$, and contrary to
expectations, it is not reduced in the underdoped materials.  This
has implications for
understanding the increased c-axis confinement in underdoped
materials.

\end{abstract}
\vspace*{-0.3 in}

It has long been thought that the high temperature superconducting
cuprates can to first order be treated as two dimensional systems
since the transport properties within the ab-plane ($CuO_{2}$ plane) and
along the c-axis are very different.  Such anisotropy shows up for
example in the resistivity ratio $\rho_{c}/\rho_{a,b}$ and their
temperature dependence $\rho_{(a,b),c}(T)$\cite{Watanabe-rho,Ando}.  Other
probes such as the optical conductivity and interlayer tunneling
measurements also indicate that the charge transport along the c-axis
is severely reduced, especially in underdoped
samples\cite{Gough,Puchkov}.  This leads to the general notion that
the physics of these materials is almost entirely two dimensional, and
can be well described by a single $CuO_{2}$ plane.  However, this picture
seems to be incompatible with the fact that the transition temperature
$T_{c}$ is closely related to the number of $CuO_{2}$ planes per unit
cell\cite{TC}, with single layer compounds of a family generically
having lower $T_{c}$'s than bilayer or trilayer compounds.
Additionally, recent experiments on the organic-chain intercalated
$Bi_{2}Sr_{2}CaCu_{2}O_{8+\delta}$ (Bi2212) showed that isolated 
$CuO_{2}$
planes do not possess superconductivity\cite{organic-chain}.  These
results highlight the importance of the c-axis couplings for the
superconductivity.  Therefore, although they have been largely
neglected so far, it appears clear that studies of the c-axis coupling
effects on the electronic structure will be critical for a clear
understanding of the high $T_{c}$ phenomenon.

A large advance was made recently by angle-resolved photoemission
(ARPES) experiments on the overdoped bilayer cuprates
$Bi_{2}Sr_{2}CaCu_{2}O_{8+\delta}$ (Bi2212) which showed for the first
time the existence of the normal state c-axis intracell
coupling\cite{bilayer-Chuang,bilayer-Feng}.  The coupling between
adjacent $CuO_{2}$ planes within the bilayer, $t_{\perp}$, was observed as
a splitting of the near Fermi energy ($E_{F}$) bands into a Bonding
(B) and Antibonding (A) set.  This effect was found to be maximal at
the $(\pi,0)$ point of the zone and to go to zero along the
$(0,0)-(\pi,\pi)$ nodal line, consistent with theoretical
predictions\cite{OKAnderson}.  The extracted coupling value ~
$t_{\perp}$ of $55 meV$ is reduced from band theory predictions by
about a factor of 3, indicating sizable correlation effects.

Considering the highly unusual physics in the underdoped regime
including in particular the pseudogap\cite{pseudogap} and the
non-Fermi liquid behavior due to some sort of increased
correlations\cite{Varma}, it is natural to wonder whether the
correlations which reduce $t_{\perp}$ are enhanced in the underdoped
regime as well.  General expectations are that they would, since (1)
the c-axis conductivity (DC\cite{Watanabe-rho} and optical
conductivity\cite{Puchkov}) is greatly reduced in the underdoped
samples, indicating greater two dimensional confinement, (2)
correlations are expected to be stronger as the parent Mott insulator
is approached, and (3) the bilayer splitting has not yet been directly
observed in the underdoped regime.  In this letter, we present the
first direct evidence that bilayer splitting exists generically in the
Bi2212 family irrespective of doping, and in fact, the energy
splitting is essentially constant as a function of doping.

We performed high resolution ARPES on
$Bi_{2}Sr_{2}CaCu_{2}O_{8+\delta}$ over a wide range of doping from
heavily overdoped (OD, $T_{c}=55K$), through optimal (OpD,
$T_{c}=91K$) and underdoped (UD, $T_{c}=78K$), where the doping levels
corresponding to different oxygen concentrations were achieved by high
temperature anneals \cite{anneal}.  The experiments were carried out
at the Advanced Light Source (ALS), Berkeley using the HERS Scienta
endstation at the undulator BL10.0.1.  Experimental details can be
found elsewhere\cite{science-CMR}.  An important aspect of the
experiments was the choice of photon energy, which we found adjusts
the relative intensities of the bonding and antibonding bands,
allowing an easier experimental deconvolution of the various features.

Figure 1 shows experimental spectra along the $(\pi,0)-(\pi,\pi)$ line
(green line of the Brillouin zone in panel (g), which also shows a
schematic of the bonding FS (red) and antibonding FS
(black))\cite{normalization}.  False color images of the energy and
momentum spectra along this cut are shown in panels (a)-(f) for the
three dopings and for two photon energies ($22eV$, $47eV$).  The
bilayer-split bands labeled A and B are seen very clearly in panel (a)
(OD, $22eV$), in the (red) energy distribution curve (EDC) in panel
(h) and (red) momentum distribution curve (MDC) in panel (i).  In the
OpD and UD samples however, the intrinsic linewidth of the individual
features is increased to a value equivalent to or even greater than
the bilayer splitting energy, effectively masking the splitting.

Additional insight can be obtained by using the complementary photon
energy of $47eV$, which we empirically found enhances the antibonding
band A relative to the bonding band B (see panels (b), (h) and (i)).
A recent first-principles calculation of the photoemission matrix
elements has independently shown this same effect\cite{Bansil}.  Aside
from the matrix element effect changing the intensity ratio of the two
bands, we find that the lineshape and energy/momentum position of the
A band is to within experimental uncertainty unchanged.  This large
matrix element modulation allows us to deconvolve the contribution
from each of the bilayer-split bands in the optimal and underdoped
samples, where the broadening of the individual features is relatively
large.

Similar to what we have observed in panels (a) and (b), the $47eV$
images of (d) and (f) show a larger concentration of spectral weight
near $E_{F}$ than do the $22eV$ images of (c) and (e).  This is seen
more clearly in the EDCs at $(\pi,0)$ shown in (j) and (l) which show
a sharper feature at $47eV$(blue) than at $22eV$(red).  We argue that
this is because the $47eV$ data shows significant weight from the A
band, while the 22 eV data additionally shows weight from the B band.
The MDCs at $E_{F}$ (panel (i), (k) and (m)) tell a similar story -
the $47eV$ data is more concentrated at $k_{y}=0.0$ corresponding to
the A band, while the $22eV$ data extends farther out in $\vec k$
space because it contains weight from the B bands as well.

Figure 2 shows the Fermi Surfaces plots extracted from the data,
including the effects of superstructure bands and some photon energy
dependences.  Panels (a) -(f) show the the locus of normalized
near-$E_{F}$ ( in $\pm 10 meV$ window) spectral weight throughout much
of the Brillouin zone for the same OD and OpD samples and photon
energies as in figure 1.  The data were taken in the normal state
($T=80K$ for OD and $100K$ for OpD and UD) and the $\vec k$ space
location is indicated by the green box in panel (g).  A few of the
additional superstructure bands possible are also indicated in panel
(g)\cite{SS}.  Modulo matrix element effects which may be significant,
the high intensity locus of this data should represent the Fermi
Surface topology\cite{Aebi}.

As in figure 1, the bilayer splitting is most clear in panel (a) of
figure 2 (OD, $22eV$) where we see the A (black) and B bands (red)
plus some of their superstructure replicas (light black and red
lines).  Also similar to figure 1, the $47eV$ data emphasizes the A
band and so panel (b) (OD, $47eV$) almost exclusively shows the A band
with a great amount of weight near $(\pi,0)$ (in fact the matrix
elements at $47eV$ favor the second zone somewhat, causing the high
weight portion to be slightly off of the $(\pi,0)$ point).  Panels
(d), (f)
show the FS topology very similar to that in (b), i.e.  they all show the A
FS topology.  Panels (c) and (e) at $22eV$ are dominated by the standard
hole-like FS corresponding to the B band.

The above data is relevant for the issue of the Fermi Surface topology
of Bi2212, which has been an intensely debated issue recently.  In
particular, there has been a discussion about whether the Fermi
Surface may sometimes (depending upon photon energy) exhibit a more
electron-like character\cite{e-FS} , or whether there is a single
unchanging hole-like piece plus superstructures, with no effects other
than matrix elements which vary as a function of photon
energy\cite{h-FS}.  Figures 1 and 2 show that both FS topologies can
be seen on any one sample depending upon the photon energy used, and
this can be clearly understood to be simply due to the bilayer
splitting effect.  We believe that this should close the debate of the
possibility of two types of Fermi Surfaces in Bi2212- they both exist.
Whether one, the other, or both are observed depends upon the matrix
elements which naturally vary with photon energies\cite{adam-FS}.

We note that the current data can not explicitly determine which side 
of $(\pi,0)$ the antibonding band crosses, as it approaches 
exceedingly close to this point (this is the reason for the shading in 
panel (g) of figures 1 and 2).  In other words, the van Hove 
singularity of the antibonding band is very near the Fermi level at 
this point.  Typical notion is that if the band crosses $E_{F}$ along 
$(0,0)-(\pi,0)$ line, the van Hove singularity is above the $E_{F}$ 
whereas if it crosses along $(\pi,0)-(\pi,\pi)$ then it is below 
$E_{F}$.  In general it had been known that the van Hove singularity 
was near $E_{F}$, with a well known parameterization being $34 meV$ 
below $E_{F}$\cite{Norman-hFS}.  While this is quite near $E_{F}$, it 
is still much larger than the temperature scale of $T_{c}$.  On the 
other hand, the current data indicates that for the overdoped sample 
the antibonding band is at binding energy $9 \pm 2meV$ at the 
$(\pi,0)$ point, which is comparable to the temperature scale of the 
problem\cite{fitting}.  This implies it should be quite active 
electronically.  The van Hove singularity of the standard 
``hole-like'' Fermi Surface (bonding component) is on the other hand 
about $100 meV$ below $E_{F}$, making it much less relevant.  This is 
important, as there have been many theoretical suggestions that a van 
Hove singularity near $E_{F}$ may be critical to understand the 
physics of these compounds\cite{Markiewicz,Abrikosov,Newns} .

The existence of the bilayer splitting and its effect on the Fermi
Surface topology is qualitatively new and was obtained directly from
the raw data with no fitting or modeling required.  This is the ideal
most trustworthy case.  There is much new quantitative information as
well, and this can be extracted through a simple fitting of the data.
Near $E_{F}$ where the dispersion is steep we fit each set of MDCs
with two Lorentzian peaks and a linear background (see for example
figures 3(a)-(d)).  Near the bottom of the bands where the dispersion
is flat we fit the EDCs with two Lorentzian peaks plus a linear
background multiplied by a Fermi cutoff (figure 3(e)-(g)).  Usually we
can only fit a portion of the spectra (positive $k_{y}$ portion in
(a), negative $k_{y}$ in (b),(c),(d)) because of contamination by
superstructure bands.  The easiest data to fit is $22eV$ OD, which
shows both A and B bands clearly and well separated (figure 3(a) and
3(e)).  However, data from the optimal and underdoped samples can also
be well fit if we tune to $47eV$ and use two bands A and B. It is much
more difficult or impossible to fit these spectra with one
peak\cite{SC-UD}.

The overall fitting results are compiled in figures 3(h) - 3(k), each
set of which corresponds to about 20 fits of the type in panels
(a)-(g).  These were each taken along a cut about $27\%$ of
the zone away from $(\pi,0)$ where the superstructure contamination
was minimal ($k_{x}=0.73\pi$ for $22eV$ data and $1.27\pi$ for $47eV$
data).  The open circles are the EDC peak centroids and the closed
circles are the MDC peak centroids.  The dashed lines are quadratic
fits to the centroids, implying a simple parabolic band
dispersion\cite{kink}.  The fact that both the MDC and EDC centroids
are very well fit by the same parabolic bands gives high confidence
for the quality of the data and fits for all sample types.

Panels (i) - (k) show the parabolic fits from all three sample types
measured under identical conditions ($47eV$, $k_{x}=1.27\pi$ cut,
normal state).  While the different doping levels give an energy
offset of the parabolas from one sample to another ($30 meV$ from OD
to OpD and another $10 meV$ from OpD to UD) due to the band filling
effect, the shape and energy splitting of the parabolic curves are
essentially unchanged with doping.  In practice, we found the shape
and splitting ($\approx 70 meV$) of the parabolic curves from the OD
sample and then overlaid this parabola (shifted in energy but with the
same splitting) with the data for the other samples.  The good
agreement is clear from the fits.

The bilayer splitting as a function of k has been theoretically
parameterized as $\Delta(\vec
k)=0.5t_{\perp}(cos(k_{x}a)-cos(k_{y}a))^{2}$, which is zero along the
node lines and maximal at the $(\pi,0)$ points\cite{OKAnderson}.
Using this formula to extrapolate the current data to $(\pi,0)$ we get
a maximum splitting of $114 meV\pm 8 meV$ or a $t_{\perp}\approx 57
meV \pm 4 meV$, obtained from the $47eV$ data.  The $22eV$ data shown
here (panel (h)) gives an estimate of $54meV$ also.  This is in excellent
agreement with our previous estimate of $55meV$ obtained on $25eV$
data\cite{bilayer-Chuang}.  Thus we conclude that $t_{\perp}$ is
independent of the incident photon energies and has at the most a very
weak dependence on the doping level.

The conclusion that $t_{\perp}$ depends weakly on the hole doping
levels is quite surprising since the normal state resistivities show
very different behavior, both in the magnitude and the temperature
dependence\cite{Watanabe-rho,Ando}.  For example, in the UD regime,
$\rho_{c}$ shows insulating behavior while $\rho_{ab}$ is still
metallic.  In contrast, both $\rho_{ab}$ and $\rho_{c}$ are metallic
in the overdoped regime.  This type of data has been widely taken to
indicate an increased two-dimensional confinement in the underdoped
samples, a phenomenology that would be consistent with other exotic
behaviors such as spin-charge separation\cite{PWAnderson} or marginal
Fermi Liquid behavior\cite{Varma}.  While coupling between the unit
cells is also relevant for the confinement, this coupling should be
much weaker than the bilayer coupling discussed here, and in fact this
coupling has not yet been directly observed.  Our data directly
implies a lack of increase in two dimensional confinement from the
strong intracell $t_{\perp}$ and hints at a lack of change from the
intercell $t_{\perp}$ effect.  We suggest that other factors such as
the increase in the scattering rate near $(\pi,0)$, the onset of the
pseudogap, or carrier fractionization should be more responsible for
these anomalies in the transport behavior.

We acknowledge sample preparation help from J. Koralek and M. Varney
and beamline support from S. Kellar, X.J. Zhou, P. Bogdanov, and Z.
Hussain.  This work was supported by the NSF Career-DMR-9985492 and
the DOE DE-FG03-00ER45809.  ALS is operated by the DOE, Office of
Basic Energy Sciences.


\begin{figure}
      \caption{(a)-(f) Energy and momentum spectra along the
      $(\pi,0)-(\pi,\pi)$ line, with a  false color intensity
      scale. (g) A Fermi Surface plot indicating the location of the data
      cuts (green lines).  Superstructure bands are not shown, and the
      shading near $(\pi,0)$ indicates the uncertainty of the FS
location.  (h),(j),(l)
      EDCs at the $(\pi,0)$ point. (I),(k),(m) MDCs at $E_{F}$. Red
curves are  $22eV$ data and blue curves
      are $47eV$ data. }
\end{figure}

\begin{figure}
	 \caption{Integrated spectral weight over $(-10meV,10meV)$ binding
	 energy window.  (a) OD,$h\nu=22eV$.  (b) OD, $47eV$. (c) OpD,
	 $22eV$. (d) OpD, $47eV$. (e) UD,
	 $22eV$.  (f) UD, $47eV$.  The shaded area of 1(g) is removed in
	 these panels.  (g) schematic FS topology with A FS (black) and B
	 FS (red), plus a few of the SS bands (lighter).}
\end{figure}

\begin{figure}
      \caption{Fitting results on the $47eV$
      data at $k_{x}=1.27\pi$ and on the $22eV$ data at $k_{x}=0.73\pi$.
      (a)-(d) Overlay of MDC at $E_{F}$ with
      fitted curves. (e): Overlay of EDC at $(0.73\pi,0.1\pi)$ at
$22eV$ with fitted
      curves. Open circles are the data, dashed lines are Lorentzian
      peaks, dotted lines are the backgrounds and solid lines are the overall
      fitting results.
      (f),(g) Overlay of EDCs at $(1.27\pi,-0.1\pi)$ with fitted curves.
      (h) -(k)
      Overlay  of the centroids from EDCs (open circles) and MDCs (closed
      circles) with
      parabolic fitting curves (dashed lines).}
 
\end{figure}

\vspace*{-0.2in}


\begin{references}
\bibitem{anneal} OD samples with a $T_{c}$ of $55K$ were
obtained by annealing as grown crystals at $T=400 ^{o}C$ under a pressurized
($\gtrsim$ 1000 atm) Oxygen atmosphere; OpD samples with a $T_{c}$ of
$91K$ were obtained by annealing at $T=700 ^{o}C$ in air and UD
samples with a $T_{c}$ of $78K$ were annealed
in a reduced Oxygen atmosphere (400 torr Argon/100ppm Oxygen) at
$T=400 ^{o}C$.
\bibitem{normalization} The EDCs used in making such plots
were obtained in the normal state and were normalized by
using high harmonic emission above $E_{F}$, which is believed to be
isotropic in momentum space.
\bibitem{SS} These are due to the 5x1 superstructure modulation along 
the b or $\Gamma - Y$ axis.
\bibitem{holeFS} the red curves are the hole-like Fermi surface
topology taken from ref \cite{Ding-PRL}. The thick red lines are the 
main Fermi surface
and the thin red lines are the Fermi surface from the lattice
distortions (superstructure bands).
\bibitem{fitting}  The curves at $k_{x}=1.0$ were fit in a similar
way to the data of figure 3.
\bibitem{SC-UD} The feature observed in UD sample at $47eV$ does show
the superconducting gap of $\approx 35meV$ when we cooled down to
$T=30K$ which indicates its relavance to superconductivity. Data is
not shown here.
\bibitem{kink} We note that in the normal state near the $(\pi,0)$
point there is no clear mass renormalization or "kink".
\end{references}
\end{document}